\documentclass[
  reprint,
  amsmath,amssymb,
  aps,
  prx,
  superscriptaddress
]{revtex4-2}
\usepackage{graphicx}
\usepackage{dcolumn}
\usepackage{bm}
\usepackage{booktabs}
\begin{document}

\title{Making Weak Interlayer Excitons Absorb via Purcell Engineering of Critical Coupling}

\author{Saroj Poudyal}
\affiliation{School of Electrical Engineering, Faculty of Engineering,
             Tel Aviv University, Tel Aviv 6997801, Israel}
\author{Anabel Atash Kahlon}
\affiliation{School of Electrical Engineering, Faculty of Engineering,
             Tel Aviv University, Tel Aviv 6997801, Israel}
\author{Michael Klein}
\affiliation{School of Electrical Engineering, Faculty of Engineering,
             Tel Aviv University, Tel Aviv 6997801, Israel}
\author{Matan Meshulam}
\affiliation{School of Electrical Engineering, Faculty of Engineering,
             Tel Aviv University, Tel Aviv 6997801, Israel}
\author{Yonatan Gershuni}
\affiliation{School of Electrical Engineering, Faculty of Engineering,
             Tel Aviv University, Tel Aviv 6997801, Israel}
\author{Thomas Poirier}
\affiliation{Tim Taylor Department of Chemical Engineering,
             Kansas State University, Manhattan, KS 66506, USA}
\author{James H. Edgar}
\affiliation{Tim Taylor Department of Chemical Engineering,
             Kansas State University, Manhattan, KS 66506, USA}

\author{Mukesh Kumar}
\affiliation{School for Engineering of Matter, Transport and Energy,
             Arizona State University, Tempe, AZ 85287, USA}
\author{Seth Ariel Tongay}
\affiliation{School for Engineering of Matter, Transport and Energy,
             Arizona State University, Tempe, AZ 85287, USA}
\author{Itai Epstein}
\affiliation{School of Electrical Engineering, Faculty of Engineering,
             Tel Aviv University, Tel Aviv 6997801, Israel}
\affiliation{Center for light-Matter Interaction,
             Tel Aviv University, Tel Aviv 6997801, Israel}
\affiliation{QuanTAU, Quantum Science and Technology Center,
             Tel Aviv University, Tel Aviv 6997801, Israel}
\begin{abstract}
Interlayer excitons (IXs) in transition metal dichalcogenide bilayers possess unique dipolar properties, providing a path for exploring correlated many-body phases, electrically tunable dipole-dipole interactions, and long-range exciton transport. However, the large electron–hole separation underlying their dipolar character also results in a significantly reduced oscillator strength, limiting their  interaction with light, and thus hindering the direct optical probing of their dynamics and many-body interactions. Here, we show that by using Purcell engineering of critical coupling, a weak interlayer transition can absorb as strongly as the dominant intralayer transition, despite its much smaller intrinsic oscillator strength. Using this approach, in an open cavity architecture we demonstrate an order of magnitude IX absorption modulation in bilayer and trilayer MoS\(_2\), with doubling of the IX to A-exciton interaction strength weight from $\sim23\%$ to $\sim53\%$. Furthermore, we demonstrate that in a tailored closed cavity architecture, the IX enhancement can reach beyond an order of magnitude, from $\sim5\%$ to $\sim83\%$ excitonic absorption, enabling IX absorption similar to intralayer excitons. An analytical description of the interaction is obtained via a closed-form transmission-line model which separates the excitonic and photonic contributions to the optical response, revealing how cavity architecture can be designed towards critical coupling. Our findings establish Purcell engineering of critical coupling as a general route to controlling and enhancing weak excitonic transitions.

\end{abstract}
\maketitle

\section{Introduction}
Interlayer excitons (IXs) in transition metal dichalcogenides (TMD) bilayers have attracted considerable
scientific interest, as possible solid-state platforms for dipolar exciton
physics and excitonic optoelectronics \cite{Sun2024DipolarHeterobilayers,Regan2022EmergingHeterobilayers,Liu2022InterlayerOptoelectronics,Du2023MoireOptoelectronics,Azimi2025}. They host correlated many-body
phases, such as exciton condensates and excitonic insulators, act as
moir\'e-trapped quantum emitters, and underpin electrically switchable
excitonic transistors, photodetectors, and emitters
\cite{Regan2022EmergingHeterobilayers,Sun2024DipolarHeterobilayers,Liu2022InterlayerOptoelectronics,Jiang2021InterlayerHeterostructures,Mak2018OpportunitiesManipulation,Sun2021ExcitonicHeterostructure,Lukman2020HighPhotodetection}.
In TMD hetero-bilayer systems, type-II band alignment is commonly formed, where the electron and hole constituting the exciton reside
in different layers, giving rise to a permanent out-of-plane
electric dipole moment and long lifetimes, reaching tens to hundreds of
nanoseconds
\cite{Rivera2015ObservationHeterostructures,Rivera2016Valley-polarizedHeterostructure,%
      Unuchek2018Room-temperatureHeterostructure}.

While this configuration brings about fascinating attributes, such as electrically tunable dipole-dipole
interactions, long-range exciton transport, and dipolar many-body
phenomena
\cite{Jauregui2019ElectricalHeterostructures,%
      Unuchek2018Room-temperatureHeterostructure,%
      Steinhoff2024Exciton-excitonHeterobilayers,Sun2021ExcitonicHeterostructure}, the spatial separation of charge carriers also yields 
weak overlap of their wavefunctions, resulting in a reduced oscillator strength about two orders of magnitude smaller than that of
intralayer excitons
\cite{Ross2017InterlayerJunction,Barre2022OpticalHeterostructures}. Consequently, IXs appear only as faint features in reflection spectroscopy, if at all, compared to their intralayer counterparts, and their direct interaction typically
requires complex techniques, such as modulated reflectance or photocurrent measurements \cite{Barre2022OpticalHeterostructures,Wietek2024NonlinearHeterobilayers}. Even in homobilayer TMDs, such as MoS$_2$, where hybridization between the valence bands of the two layers yields IXs with a finite oscillator strength and enables their resonant interaction with light, they only reach oscillator strengths of about $25\%$ of the intralayer excitons
\cite{Gerber2019InterlayerTemperature,Leisgang2020GiantMoS2,Zhao2022InterlayerMoS2}.
Despite their weak oscillator strength, hybridized IXs in bilayer MoS$_2$ have recently been shown to host electrically tunable
Coulomb interactions, ultrafast formation and relaxation dynamics, to serve as sensitive optical
probes of correlated electronic phases, such as interlayer electron coherence, and to form dipolar polaritons in microcavities
\cite{Liu2025OpticalSemiconductor, Lopriore2025ElectricallyBilayer,Leisgang2020GiantMoS2,Peimyoo2021ElectricalMoS2,Datta2022HighlyMoS2,Sponfeldner2022CapacitivelyMoS2,Lorchat2021ExcitonsResponse,Louca2023InterspeciesHomobilayers,Genco2025FemtosecondSemiconductors,Kovalchuk2025RevealingGating}. Nevertheless, their direct interaction with light remains weak due to the much smaller oscillator strength, compared to intralayer excitons, corresponding to absorption of about $\sim2-4\%$ \cite{Li2014MeasurementE2}. Thus, controlling their effective oscillator strength is
essential to both accessing and harnessing their properties. 

Previous efforts to enhance the photoluminescence of IXs in type-II heterobilayer systems have included tunable fiber microcavities,
whispering-gallery-mode resonators, high-index photonic crystal nanocavities, and chirped distributed Bragg reflector microcavities
\cite{Forg2019Cavity-controlHeterostructures,Han2025Yield, Khelifa2020CouplingHeterostructures, Rivera2020CouplingDichalcogenides,Qian2024LasingNanocavity, Palekar2021TunableScenarios,Lopriore2025EnhancingEffect,Rahaman2023TailoringRegime}.
However, all of these have addressed IX emission processes, and a route to directly controlling the light-IX interaction via the manipulation of their small
oscillator strength has remained elusive.

In this work, we demonstrate how the effective oscillator strength of IXs in bilayer MoS\(_2\), and consequently their direct interaction with light, can be significantly and selectively enhanced through Purcell engineering of critical coupling, as it favors  weakly coupled resonances. Using this approach, in open cavity architecture we demonstrate an order of magnitude IX absorption modulation in bilayer and trilayer MoS\(_2\), with doubling of the IX-to-intralayer exciton interaction strength weight from $\sim23\%$ to $\sim53\%$. Furthermore, we demonstrate that in a tailored closed cavity architecture, the IX enhancement can reach beyond an order of magnitude, from $\sim5\%$ to $\sim83\%$, raising the IX to A-exciton weight to $\sim 99\%$. To analyze and predict the light-IX interaction in such architectures, we develop a closed-form transmission-line model (TLM), which separates the cavity and excitonic contributions to their interaction with light, enabling an overall selective enhancement of weakly absorbing IXs via Purcell-engineering of critical coupling.

\section{Characterization of IX\lowercase{s} in bilayer M\lowercase{o}S$_2$}
\label{sec:char}
Fundamentally, resonant absorption of light in an optical system is
governed by the competition between its radiative and nonradiative decay
rates, which for excitons in TMDs enter the optical response through
the in-plane Lorentzian susceptibility  \cite{Epstein2020Near-unityCavity}
\begin{equation}
  \chi_{\parallel}(\omega)
  = \chi_{\mathrm{bg}}
    - \sum_j
      \frac{c}{\omega_j d_0}
      \frac{\gamma_{r,0}^{(j)}}
           {(\omega - \omega_j) + i\,\gamma_{nr}^{(j)}/2},
  \label{eq:chi}
\end{equation}
where $\chi_{\mathrm{bg}}$ is the background susceptibility, $\omega_j$
the frequency of the $j$th exciton resonance, and $d_0 = 1.4~\mathrm{nm}$ is
the bilayer's thickness. The vacuum radiative decay rate, $\gamma_{r,0}$, is an intrinsic excitonic property,  which directly
relates to its oscillator strength and thus to its
interaction with light \cite{Barre2022OpticalHeterostructures,%
Horng2020PerfectCrystal}, and $\gamma_{nr}$ is the nonradiative rate corresponding to all population loss channels that do not lead to photon emission. When these rates are balanced, the system is said to be critically coupled, and the incident power is maximally absorbed by the resonance \cite{Lee2023AchievingNesting,Rogers2020CoherentExcitons,Horng2020PerfectCrystal,Epstein2020Near-unityCavity,Li2023AModulator}. However, it is important to note that these decay rates are intrinsic physical properties of the system, and thus the ability to control and manipulate their balance to achieve critical coupling is generally non-trivial, and even unattainable in most cases.

The radiative rate of an emitter, such as an exciton, can be changed by manipulating its local photonic density of states. This phenomenon is known as the Purcell effect \cite{Purcell1946SpontaneousFrequencies,Drexhage1970InfluenceTime}, 
and the exciton radiates at an effective radiative rate, $\gamma_r =\zeta \gamma_{r,0}$, where $\zeta$ is the Purcell factor. The dissipation, or absorption, becomes 
maximal when the two decay rates are matched, i.e. 
$\gamma_{r} = \gamma_{nr}$, and the system is said to be at critical coupling \cite{Hinney2024EfficientCoupling,RaDi2020VirtualCoupling,%
Rogers2020CoherentExcitons,Horng2020PerfectCrystal,%
Epstein2020Near-unityCavity}. The intrinsic weight, $W_0 = \gamma_{r,0}/\gamma_{nr}$, therefore measures how far the system is from critical coupling, which is reached when  $\zeta \, W_0 =1$.

\begin{figure}
  \centering
  \includegraphics[width=1\linewidth]{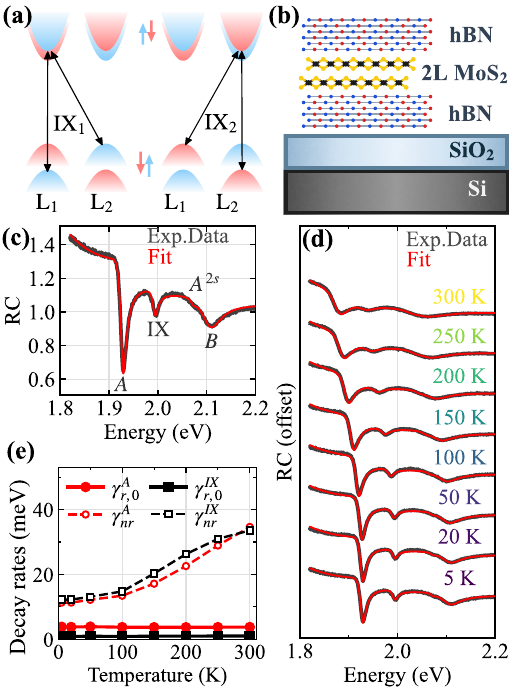}
  \caption{Characterization of IX in bilayer MoS\(_2\).
    (a)~Schematic band structure of bilayer MoS$_2$ near the $K$ point
    illustrating the formation of IXs. Blue and red colors denote
    spin-up and spin-down states respectively, while L1 and L2
    correspond to the two layers of the bilayer.  (b)~Schematic diagram of device geometry for characterization of the IX in which bilayer MoS$_2$ is encapsulated by hBN on a SiO$_2$/Si substrate. (c) ~RC spectrum (black dotted line) of hBN-encapsulated bilayer
    MoS$_2$ measured at $T\!=\!5\,\mathrm{K}$ on a SiO$_2$/Si
    reference substrate, together with the TLM fit (red line). The
    four resonances corresponding to the A exciton, the IX, the
    Rydberg state of the A exciton, and the B exciton are marked. 
    (d)~Temperature dependent RC measurements from $T\!=\!5\,\mathrm{K}$ to room temperature. (e)~Extracted radiative and nonradiative decay rates of the intralayer A exciton and the IX as a function of temperature. }
  \label{fig:IXcharacterization}
\end{figure}

The physical properties of IXs in bilayer MoS\(_2\) stem from strong hybridization of the two layers \cite{Gerber2019InterlayerTemperature, Deilmann2018InterlayerMaterials,Das2019LayerDichalcogenides}. The strongly-hybridized valence bands result in a delocalization of the holes over both layers, while the electron remains localized in either layer~1 or layer~2, as shown schematically in Fig.~\ref{fig:IXcharacterization}(a), and the two resulting configurations are energetically degenerate (black arrows in Fig.~\ref{fig:IXcharacterization}(a)) \cite{Leisgang2020GiantMoS2}. To assess the critical coupling aspects of IXs in bilayer MoS$_2$, it is first required to characterize their decay rates, which to the best of our knowledge has not been previously reported. For this purpose, we fabricated high-quality hBN-encapsulated bilayer MoS\(_2\) \cite{Raja2017CoulombMaterials,Itzhak2025ExcitonSemiconductors} ,  on a SiO\(_2\)/Si substrate (Fig.~\ref{fig:IXcharacterization}(b)) (see Appendix~\ref{sec:sample} for fabrication details). Figure~\ref{fig:IXcharacterization}(c) presents the reflection-contrast spectrum, $RC={R}/{R_0}$, from an exemplary device (black dotted line), measured at $T\!=\!5\,\mathrm{K}$, where $R$ and $R_0$ are the reflectance measured with and without the bilayer MoS$_2$, respectively (see Appendix~\ref{sec:measurement} for the measurement details). The A and B intralayer excitonic resonances can be clearly seen \cite{Wang2018Colloquium:Dichalcogenides}, with the IX resonance located between them as expected, and the ${A^{2s}}$ corresponding to the $2s$ Rydberg state of the A exciton \cite{Gerber2019InterlayerTemperature}. It can be seen in Fig.~\ref{fig:IXcharacterization}(c) that the reflection spectrum is dominated by the intralayer A exciton, while the IX appears as a comparatively weaker resonance, reflecting its reduced oscillator strength.

The fitted spectrum at $T\!=\!5\,\mathrm{K}$, obtained via the TLM by modeling the bilayer by the set of Lorentzian oscillators in Eq.~\eqref{eq:chi}, is shown in Fig.~\ref{fig:IXcharacterization}(c) (red line) \cite{PhysRevLett.134.196903,https://doi.org/10.1002/adom.202501712, PhysRevB.111.L041302,Kahlon2025ImportanceHeterostructures}, and the corresponding extracted decay rates are shown in Fig.~\ref{fig:IXcharacterization}(e)  (see Table~\ref{tab:decaytable} for the values). As the oscillator strength is proportional to the radiative decay rate, we find the intrinsic oscillator strength of the IX to be $\approx 26 \%$ of the A exciton, consistent with prior reports \cite{Gerber2019InterlayerTemperature, Peimyoo2021ElectricalMoS2,Datta2022HighlyMoS2}, and the corresponding intrinsic weights in bilayer MoS\(_2\) are $W_{0,A} \approx 0.341$ and $W_{0,IX} \approx 0.081$, for the A exciton and IX, respectively. Both resonances are therefore found to be far from critical coupling, with the IX being much further away. 

The complete temperature-dependent fits of RC are presented in Fig.~\ref{fig:IXcharacterization}(d) together with the extracted values of $\gamma_{r,0}$ and $\gamma_{nr}$ in Fig.~\ref{fig:IXcharacterization}(e). It can be seen that the $\gamma_{r,0}$ are nearly constant with temperature, as expected \cite{Epstein2020Near-unityCavity,Meshulam2026Temperature-DependentTMDs}, so the oscillator strength ratio of IX and A hardly changes with temperature. The $\gamma_{nr}$, in contrast, increases with temperature, as expected \cite{Guarneri2024Temperature-DependentElements,Nauman2025DynamicMetasurfaces}, moving both resonances further away from critical coupling. For completeness, we have measured a second reference device, giving similar results ( Appendix~\ref{sec:additionaldevices}).

\section{Analytical analysis of cavity Purcell engineering via the Transmission-line model}
\label{sec:tlm_main}
\begin{figure}
  \centering
  \includegraphics[width=\linewidth]{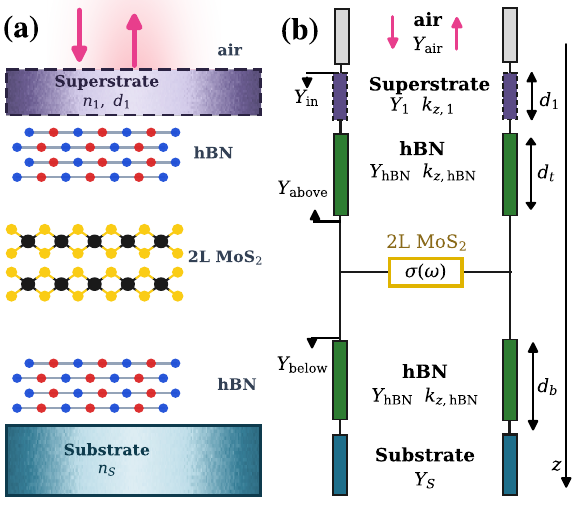}
  \caption{Transmission-line model of the general planar van der Waals
    heterostructure.
    (a)~Physical stack,
    air\,/\,superstrate\,($n_1,d_1$)\,/\,hBN\,($d_t$)\,/\,bilayer
    MoS$_2$\,/\,hBN\,($d_b$)\,/\,substrate\,($n_S$). The superstrate is
    dashed as it may be absent, a thin metal film, or a Bragg stack.
    (b)~Equivalent circuit, each bulk layer a line section of admittance
    $Y_j = n_j\varepsilon_0 c$ and propagation constant $k_{z,j}$, and
    the bilayer a shunt admittance $\sigma(\omega)$ bridging the line at
    its own plane. Arrows give the direction of view.
    $Y_{\mathrm{above}}$ and $Y_{\mathrm{below}}$ are the admittances
    seen looking up and down from the bilayer plane, and together with
    $\widetilde{Y}_{\mathrm{above}}$ they fix the optical response of any
    geometry treated here (Table~\ref{tab:cases}).}
  \label{fig:tlmgeneralfig}
\end{figure}

Since $\gamma_{nr}$ is already close to its physical limit for an hBN-encapsulated bilayer and can only rise with temperature \cite{Selig2016ExcitonicDichalcogenides,Katsch2020Exciton-Scattering-InducedSemiconductors,Gupta2019FundamentalDichalcogenides}, we focus on the van der Waals heterostructure cavity architecture
\cite{Epstein2020Near-unityCavity} to Purcell-engineer the IX effective oscillator strength towards critical coupling. 
To predict the Purcell factor analytically, we use TLM, where each layer becomes a circuit element and the whole structure becomes
a one-dimensional network \cite{Eini2025TransmissionHeterostructures,Kahlon2025ImportanceHeterostructures}.
Each layer in the air\,/\,superstrate\,/\,hBN\,($d_t$)\,/\,bilayer
MoS$_2$\,/\,hBN\,($d_b$)\,/\,substrate heterostructure is modelled as a line section of characteristic admittance
$Y_j = n_j\varepsilon_0 c$, where $n_j$ is its refractive index (Fig.~\ref{fig:tlmgeneralfig}(a)). The
bilayer is much thinner than the wavelength, and enters as a shunt conductivity
\begin{equation}
  \sigma(\omega) = -i\omega\varepsilon_0
    \bigl(\chi_{\parallel}(\omega)
          -\chi_{\mathrm{sur},\parallel}(\omega)\bigr) d_0 ,
  \label{eq:sigma}
\end{equation}
which bridges the line at the hBN/hBN plane
(Fig.~\ref{fig:tlmgeneralfig}(b)), where $\chi_{\parallel}$ and
$\chi_{\mathrm{sur},\parallel}$ are the susceptibilities of bilayer
MoS\(_2\) (Eq.~\eqref{eq:chi}) and of the surrounding hBN,
respectively. 

We use the TLM in an analytical form which separates the bilayer from the cavity, in which the measured RC takes the form:
\begin{equation}
  \frac{R}{R_0} = \left|1-Z\right|^2 ,
  \qquad
  Z(\omega) = \frac{P+Q}{\dfrac{1}{\sigma}+P} ,
  \label{eq:master}
\end{equation}
where the bilayer physics resides within $\sigma$ and all of the
photonic physics in the two impedances
$P = (Y_{\mathrm{above}}+Y_{\mathrm{below}})^{-1}$ and
$Q = (\widetilde{Y}_{\mathrm{above}}-Y_{\mathrm{below}})^{-1}$, both 
evaluated at the bilayer plane (Appendix \ref{sec:tlm}). Here $Y_{\mathrm{below}}$ and
$Y_{\mathrm{above}}$ are the admittances below
and above the bilayer evaluated at that plane. $\widetilde{Y}_{\mathrm{above}}$ is $Y_{\mathrm{above}}$ with every
propagation phase reversed, which is the admittance of the incoming wave at that plane.

Inserting the Lorentzian susceptibility of Eq.~\eqref{eq:chi} into
Eq.~\eqref{eq:master} shows that each excitonic resonance keeps its Lorentzian
form but acquires a cavity-dressed linewidth
\begin{equation}
  \Gamma_T = \gamma_{nr} + \zeta(\omega)\,\gamma_{r,0},
  \label{eq:GammaT}
\end{equation}
with the Purcell factor $\zeta = 2\varepsilon_0 c\,\mathrm{Re}\,P'$. Here $\frac{1}{P'} = \frac{1}{P} +\sigma_{bg}$ is the cavity admittance screened by the
 background conductivity $\sigma_{bg}$ of the bilayer (Appendix \ref{sec:tlm}). In terms of the
two interfaces and the two hBN phases this reads
\begin{equation}
  \zeta(\omega)
  =
  \mathrm{Re}\!\left[
  \frac{\mathcal{F}(\omega)}
  {n_H+\frac{\sigma_{\mathrm{bg}}}{2\varepsilon_0 c}\mathcal{F}(\omega)}
  \right],
  \label{eq:zeta}
\end{equation}
where $n_H$ is the hBN refractive index and $\mathcal{F}(\omega)$ is the cavity function
\begin{equation}
  \mathcal{F}(\omega) =
  \frac{(1+r_b(\omega))(1+r_t(\omega))}{1 - r_b(\omega)\,r_t(\omega)}.
  \label{eq:calF}
\end{equation}
Here $r_t(\omega)$ combines the reflection coefficient of the interface between
the top hBN and the medium above it with the round-trip phase across the top hBN of thickness $d_t$. Likewise $r_b(\omega)$
combines the reflection coefficient of the hBN/substrate interface with the round trip phase across the bottom hBN of
thickness $d_b$. The whole photonic environment is thus reduced to 
these two reflectors, and each geometry in this work follows by
evaluating them for the corresponding stack (Table~\ref{tab:cases}).
 
Physically, $\mathcal{F}$ measures the field that the two reflectors
build up at the bilayer. When the hBN thicknesses place the
bilayer at an antinode of the standing wave, $\mathcal{F}$ is maximal,
and the exciton radiates into an enhanced electromagnetic environment, so
its $\gamma_{r}$ rises. At a node the field is suppressed and so is the $\gamma_{r}$. The cavity changes only the $\gamma_{r}$, since the
nonradiative rate is set by internal scattering channels.
The cavity therefore moves the balance $\zeta\gamma_{r,0}/\gamma_{nr}$ that
governs critical coupling without adding dissipation to the exciton
itself.
 
This gives a direct design rule. The largest and smallest values of $\zeta$ over all hBN
thicknesses are set by the moduli of the two reflection coefficients $|r_b|$ and $|r_t|$
alone (Appendix~\ref{sec:bounds}),
\begin{equation}
\begin{split}
  \zeta_{\max} = \frac{(1+|r_b|)(1+|r_t|)}{n_{H}(1-|r_b|\,|r_t|)},\\[6pt]
  \zeta_{\min} = \frac{(1-|r_b|)(1-|r_t|)}{n_{H}(1-|r_b|\,|r_t|)}.
  \label{eq:bound_main}
\end{split}
\end{equation}
The reflectors set the accessible range, and the two hBN
thicknesses select a point within it continuously. A single material
stack thus provides a whole family of photonic environments,
from strong enhancement to strong suppression, chosen purely by
geometry. Moreover, the peak absorption increases monotonically with $\zeta$ as long as the resonance is below critical coupling (Appendix~\ref{sec:monotone}), so the geometry
that maximizes the Purcell factor  also maximizes the absorption. Since $\zeta$ contains no decay rates, this geometry follows from the resonance energy and background susceptibility alone.

\section{Open cavities}
\label{sec:opencavities}

\begin{figure}
  \centering
  \includegraphics[width=\linewidth]{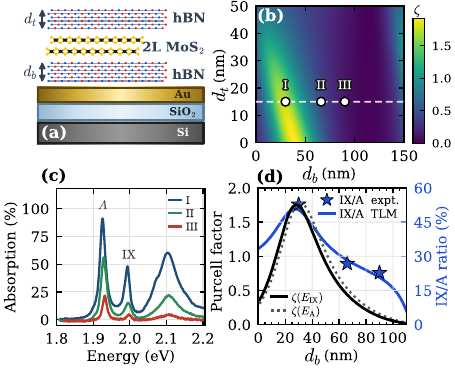}
  \caption{
    Experimental realization of open cavity.
    (a)~Device schematic:
    air\,/\,hBN($d_t$)\,/\,bilayer
    MoS$_2$\,/\,hBN($d_b$)\,/\,Au\,/\,SiO$_2$\,/\,Si.  Varying $d_b$ at
    fixed $d_t$ tunes the bilayer from a field antinode
    ($d_b\approx 30~\mathrm{nm}$, device I) to near a node
    ($d_b\approx 90~\mathrm{nm}$, device III).
    (b)~Two-dimensional map of the TLM Purcell factor $\zeta$ versus
    $d_t$ and $d_b$ at the IX energy.  The three fabricated geometries are
    marked. (c)~Absorption spectra $A=1-R/R_0$ at $T\!=\!5\,\mathrm{K}$ for device I
    ($15/30~\mathrm{nm}$), device II ($15/66~\mathrm{nm}$), and device III
    ($15/90~\mathrm{nm}$), with the A-exciton and IX peaks labeled.
    (d)~TLM Purcell factor $\zeta$ evaluated at $E_{\mathrm{IX}}$
    (solid) and $E_{\mathrm{A}}$ (dotted) versus bottom hBN thickness
    $d_b$ at fixed $d_t=15$~nm (left axis, black).  IX-to-A
    peak-absorption ratio from the  TLM (solid line)
    and from experiment (stars) for device I, II, and III (right axis,
    blue).}
  \label{fig:fig_sm}
\end{figure}

As a first step we use a highly reflective gold mirror as the substrate to control $|r_b|$, where the bilayer is encapsulated between a
top hBN of thickness $d_t$ and a bottom hBN of thickness $d_b$ (Fig.~\ref{fig:fig_sm}(a)). Varying the hBN thicknesses tunes $\zeta$ continuously
between its maximum and minimum values, as shown in Fig.~\ref{fig:fig_sm}(b), and we select three example geometries, all with
$d_t=15~\mathrm{nm}$, and with different $d_b=30$, $66$, and $90~\mathrm{nm}$
(devices I, II, and III respectively). At the energy of the IX these span more than an order of
magnitude in $\zeta$ giving $\zeta\approx 1.74$,
$0.44$, and $0.12$ (Fig.~\ref{fig:zeta_SM_energy}). 

Figure~\ref{fig:fig_sm}(c) shows the measured absorption, $A = 1 - R/R_0$
 (since the transmission through the gold is zero \cite{Epstein2020Near-unityCavity,Horng2020PerfectCrystal,%
Lee2023AchievingNesting}),  at $T\!=\!5\,\mathrm{K}$ for the three
devices. It can be seen that following the increase in $\zeta$ from device III to device I, the IX absorption rises from $\sim 5\%$  in device III to $\sim 48\%$ in device I, roughly an order of magnitude enhancement obtained merely by moving the bilayer from the cavity's node to antinode. Similarly, the A exciton absorption rises from $\sim 22\%$ to  $\sim 91\%$. These enhancements are observed all the way up to room temperature (Fig.~\ref{fig:TD_SM}), and a validation of the enhanced effective radiative rate of the three samples can be clearly seen as brighter IX emission (Fig.~\ref{fig:pl}).

The key result accompanying the enhanced absorption is its selectivity, as the cavity favors the IX over A exciton. This can be understood by noting that both resonances experience a similar Purcell factor (black curves in
Fig.~\ref{fig:fig_sm}(d) and Fig.~\ref{fig:zeta_SM_energy}), so one would naively expect their absorptions
to scale together and their ratio to stay fixed across the three
devices. However, the measured IX-to-A absorption ratio rises from $\approx 23\%$ in device III to $\approx 53\%$ in device I (blue stars in Fig.~\ref{fig:fig_sm}(d))  showing that the same cavity enhances the weaker resonance more strongly. These results agree well with the TLM prediction
(solid blue curve).

\begin{figure}
    \centering
    \includegraphics[width=1\linewidth]{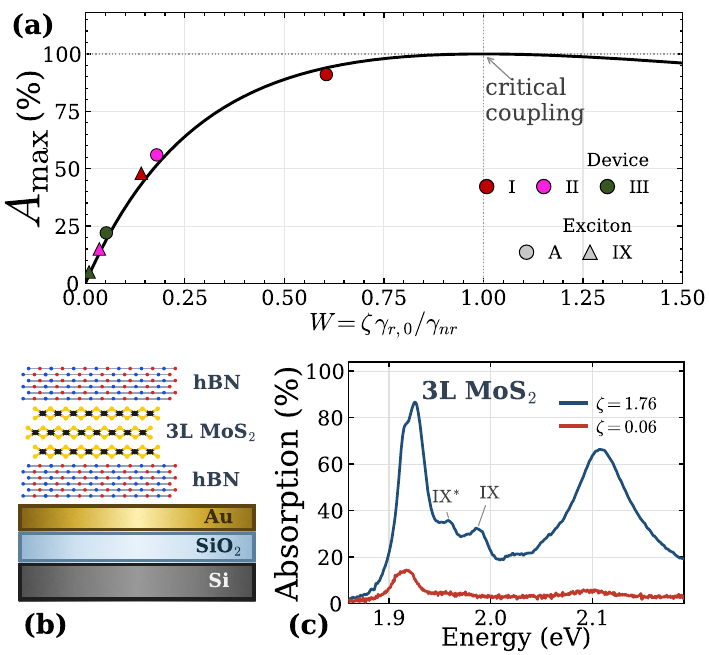}
    \caption{%
    Coupling dependent enhancement and transfer of the optimized cavity.
    (a)~Universal absorption curve $A_{\max}=4W/(1+W)^{2}$ against
    $W=\zeta\gamma_{r,0}/\gamma_{nr}$.
    The abscissa is the TLM value and the ordinate the measured peak
    absorption.
    All six operating points lie on the undercoupled branch, with the IX
    where the curve is steep and the A exciton where it has already
    flattened.
    (b)~ Schematic of the encapsulated trilayer MoS\(_2\)
    devices.
    (c)~Trilayer absorption at $T=5$~K for the antinode geometry with
    $\zeta=1.76$ and the node geometry with $\zeta=0.06$.
    The interlayer excitons are marked.}
    \label{fig:transfer}
\end{figure}

To understand the overall behavior, we use the TLM to find the peak absorption of each exciton at its resonance energy, which is well approximated in open cavities by the 
standard impedance-matching form
\cite{RaDi2020VirtualCoupling,Horng2020PerfectCrystal}
\begin{equation}
  A_{\max} = \frac{4\,W}{(1 + W)^2},
  \qquad
  W \equiv \frac{\zeta\,\gamma_{r,0}}{\gamma_{nr}}
  \label{eq:Amax_perfect}
\end{equation}
(Appendix~\ref{sec:resonance}). The peak absorption thus depends on
the cavity and excitons via $W$. The analysis is presented in Fig.~\ref{fig:transfer}(a), showing that $A_{\max}$ rises with $W$, reaching unity at critical coupling $W=1$ and falls beyond it. All measured resonances lie on the rising branch, however, since $W_{0,IX} \sim W_{0,A}/4$, the same $\zeta$ places IX and A in different regions of the curve. It can be seen in the change from device III to device I that a similar increase in $\zeta$ raises the IX absorption by 9.6 times, whereas the A exciton absorption increases only by 4.1 times. This stems from the slope of the curve, which remains steep and nearly linear at low $W$ of the IX but flattens as the A exciton approaches critical coupling. Thus, the same increase in Purcell factor selectively enhances the absorption, with a much greater preference for weakly coupled oscillators.

Since $\zeta$ does not depend on the decay rates of the excitons, the cavity can be designed using the resonance energy and the background susceptibility $\chi_{bg}$ of the sheet. To demonstrate this generality to an even more weakly coupled resonance, we apply it onto IXs in trilayer MoS$_2$ (Fig ~\ref{fig:transfer}(b)), which hosts two types of hybridized interlayer excitons, IX and IX$^{*}$ \cite{Leisgang2020GiantMoS2} (Appendix~\ref{sec:trilayerIX}). Both lie spectrally close to the bilayer IX \cite{Gerber2019InterlayerTemperature}, where $\zeta$ changes slowly, providing similar enhancement despite the different decay rates and sheet thickness. Figure ~\ref{fig:transfer}(c) presents the spectral absorption of the trilayer IXs at the cavity node and antinode, both exhibiting close to an order of magnitude enhancement, from $\sim 4 \%$ absorption to $\sim 35 \%$ absorption.

\section{Closed cavities}
\label{sec:closecavity}

\begin{figure}
  \centering
  \includegraphics[width=\linewidth]{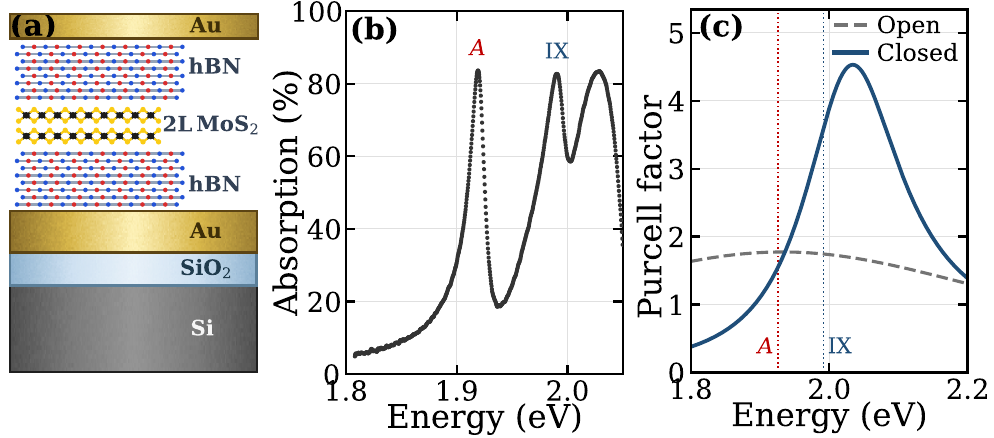}
  \caption{%
    Experimental realization of the closed cavity.
    (a)~Device schematic of
    air\,/\,Au($d_g{=}25$~nm)\,/\,hBN\,/\,bilayer
    MoS$_2$\,/\,hBN\,/\,Au.
    (b)~Absorption of bilayer MoS\(_2\) in closed cavity. The A exciton and the IX are marked. The peak near 2.03~eV is the Fabry-P\'erot mode of the cavity (see Fig.~\ref{fig:dm_fits}).
    (c)~Purcell factor $\zeta(E)$ for closed cavity
    (solid) and for the antinode open cavity device~I (dashed), with the
    A exciton and the IX marked. The open cavity is nearly flat
    across the excitonic window and therefore presents a similar photonic
    environment to the A exciton and the IX, whereas the closed cavity varies by
    nearly a factor of three and gives the IX a Purcell factor $\approx 2.3$
    times that of the A exciton.}
  \label{fig:figdm}
\end{figure}

Since the accessible Purcell factor widens with the reflectance of the top and bottom interfaces, and the bottom interface ($|r_b|$) is already close to
unity with the gold substrate, we now increase the reflectivity of the
top interface ($|r_t|$) by adding a top gold mirror (Fig.~\ref{fig:figdm}(a)). We
choose its thickness  $d_g = 25~\mathrm{nm}$ to balance transmission through the gold
film against its reflectance. The dependence of the maximum obtainable
Purcell factor and of the gold transmittance on $d_g$ is given in
Fig.~\ref{fig:dgdep}.
 
Beyond raising the accessible range, the added mirror closes the
structure into a Fabry-P\'erot cavity. The cavity resonance narrows and
becomes strongly dispersive, so that $\zeta$ varies appreciably across
the excitonic spectrum. Each exciton then experiences a different
Purcell factor, and the cavity itself becomes spectrally selective. This
is in contrast with the open cavities, where all resonances
shared a similar $\zeta$. The spectral position of the maximum of
$\zeta$ is set by the Fabry-P\'erot condition and can therefore be
placed on any chosen resonance by tuning the hBN thicknesses.
 
Figure~\ref{fig:figdm}(b) shows the measured absorption of a closed cavity with $d_t = 48~\mathrm{nm}$ and $d_b = 23~\mathrm{nm}$ at $T\!=\!5\,\mathrm{K}$. It can be clearly seen that the IX absorption reaches $83\%$, nearly equal to that of the A exciton at $84\%$. This raises the IX to A absorption ratio to $99\%$, compared with $53\%$ in the best open cavity. Thus, despite its four times smaller oscillator strength, the IX absorbs light as strongly as the intralayer exciton. This behavior originates from the spectral selectivity of the closed cavity (Fig.~\ref{fig:figdm}(c)), whereas the open cavity presents a nearly flat $\zeta$ to all resonances, in the closed cavity $\zeta$ peaks close to the IX energy and gives the IX a Purcell factor $2.3$ times larger than that of the A exciton. Selective Purcell enhancement thus adds to coupling dependent enhancement and compensates for the weak oscillator strength of the IX.

\section{Conclusion}
In conclusion, we have established Purcell engineering of critical coupling as
a general and selective route to controlling the effective oscillator strength of weakly
coupled excitons. The same approach can enhance other weakly coupled excitonic transitions
whose oscillator strengths are otherwise too small for direct optical access.
The closed-form transmission-line model provides a direct design rule that
carries over to lossless distributed Bragg reflectors and other
high-reflectivity mirrors, offering a practical path toward full critical
coupling. Combining this static photonic control with the electrical and
mechanical tunability of the interlayer-intralayer hybridization would make the position of each resonance on the absorption curve continuously tunable within a single
device. This opens new avenues for dynamically controlling the light-matter
interaction of weak oscillators in van der Waals heterostructures.

\begin{acknowledgments}
I.E. acknowledges support of the Israeli Science Foundation personal Grant No. 865/24, the Ministry of Science and Technology, via Grant 0005757, and the European Union (ERC, TOP-BLG, Project No. 101078192). J.H.E. acknowledges support from the Office of Naval Research under Award No. N00014-20-1-2474 for hBN crystal growth. S.A.T. acknowledges partial support from DOE-SC0020653 (excitonic metrology of vdW systems) and support from Applied Materials Inc., Lawrence Semiconductors Inc, CBET 2330110 (environmental tests of vdW materials), DMR 2534832 (Magnetic impurity testing), Army Research Office W911NF2510045, NSF-ECCS 2614938 (electronic topological testing), and CMMI 2533623 (plasma enhanced deposition).
\end{acknowledgments}

\appendix
\setcounter{figure}{0}
\renewcommand{\thefigure}{S\arabic{figure}}
\setcounter{table}{0}
\renewcommand{\thetable}{S\Roman{table}}

\section{Experimental methods}
\label{sec:methods}

\subsection{Material synthesis}

\noindent\textbf{MoS$_2$ Crystal Growth.} Bulk MoS$_2$ crystals were obtained from elemental Mo and S of 6N purity. Prior to use the Mo was subjected to double RF zone refining so that residual metallic and oxide impurities would be removed. Stoichiometric quantities of the two elements were weighed out in a 1 to 2 metal to chalcogen ratio inside a glovebox and were loaded into a quartz tube of 1 inch diameter. The tube was evacuated to a base pressure near $10^{-6}$ Torr and flame-sealed and the mixture was then pre-reacted at 650 degrees Celsius over a period of 24 hours in order to form the MoS$_2$ charge. The resulting powder was subsequently loaded into a fresh quartz ampoule together with iodine at a concentration of roughly 5 mg per cm$^3$ which served as the transport agent. This ampoule was likewise evacuated to approximately $10^{-6}$ Torr and sealed and was inserted into a two-zone furnace whose hot end was held at 1050 degrees Celsius while the cold end was maintained at 950 degrees Celsius. Chemical vapor transport was allowed to proceed under this thermal gradient over an interval of 7 to 10 days. The furnace was then returned to room temperature and MoS$_2$ single crystals of high quality were retrieved from the growth zone.

\subsection{Sample fabrication}
\label{sec:sample}
Bilayer and trilayer MoS$_2$ flakes in the natural 2H stacking were
mechanically exfoliated from bulk crystals and identified by optical
contrast.  Selected flakes were encapsulated between top and
bottom hexagonal boron nitride (hBN) layers using a dry,
polymer-assisted transfer technique. The   hBN thicknesses were
measured by atomic force microscopy (AFM).
Three types of heterostructures were fabricated:
\begin{enumerate}
  \item \emph{Reference devices.} An hBN-encapsulated bilayer on a
        SiO$_2$/Si substrate, used to extract the
        intrinsic excitonic parameters without a cavity.
  \item \emph{Open cavities.} The hBN
        encapsulated bilayer and trilayer were transferred onto a 100 nm gold mirror deposited by electron beam evaporation on Si/SiO\(_2\) substrate. The hBN serves here as both the cavity dielectric and the encapsulation. The bilayer
        devices differ only in the bottom hBN thickness and share the same top hBN thickness of $ 15~\mathrm{nm}$. Device I has
        $d_b = 30~\mathrm{nm}$. Device II has $d_b = 66~\mathrm{nm}$.
        Device III has $d_b = 90~\mathrm{nm}$.  Devices I and III share a
        single sample. The trilayer devices use the same geometry at
        $(d_t,d_b) = (15,30)~\mathrm{nm}$ and $(15,100)~\mathrm{nm}$.
  \item \emph{Closed cavity.} A partially transmitting
        $25~\mathrm{nm}$ Au layer was deposited on top of the
        encapsulated heterostructure by electron beam evaporation, completing a Fabry-P\'erot cavity.
\end{enumerate}

\subsection{Optical measurements}
\label{sec:measurement}

Reflection measurements were performed by focusing broadband light from
a supercontinuum source onto the heterostructure through a 40 $\times$ microscope objective (NA = 0.6). The reflected signal was collected by the same objective. The signal $R$ was acquired from a
region containing the complete hBN/bilayer/hBN structure, while the
reference $R_0$ was acquired from an adjacent region with the same
photonic stack but without MoS$_2$. Then
$RC = R/R_0$.
Photoluminescence (PL) was measured in the same setup under $532~\mathrm{nm}$
continuous-wave excitation at $1~\mu\mathrm{W}$. All spectra were
acquired at $T= 5~\mathrm{K}$ in a closed-cycle cryostat unless otherwise
stated.
\subsection{Optical constants}
\label{sec:optconst}
All transmission-line calculations use fixed optical
constants. The hBN layers are taken as lossless and non-dispersive with
refractive index $n_{H} = 2.204$ \cite{Lee2019RefractiveNear-Infrared}.
The SiO$_2$ layer of the reference substrate is
likewise non-dispersive, with $\varepsilon_{\mathrm{SiO_2}} = 2.13$
($n = 1.460$) \cite{Malitson1965InterspecimenSilica}, and the underlying
silicon is described by the tabulated constants of
Ref.~\cite{Aspnes1983DielectricEV}. The dielectric function of the gold
is taken from
Refs.~\cite{Etchegoin2006AnGold,Etchegoin2007Erratum:2006}.


\section{Reference devices }

Table \ref{tab:decaytable} lists the excitonic parameters of the reference device of Fig. \ref{fig:IXcharacterization}
\begin{table}
  \centering
  \caption{Excitonic parameters at $T = 5~\mathrm{K}$ extracted from the
           SiO$_2$/Si reference device with $\chi_{\mathrm{bg}} = 18.2$.}
  \begin{ruledtabular}
  \begin{tabular}{ldddd}
    & \multicolumn{1}{c}{A}
    & \multicolumn{1}{c}{IX}
    & \multicolumn{1}{c}{A$^{2s}$}
    & \multicolumn{1}{c}{B} \\
    \hline
    $E_0$ (eV)            & 1.926 & 1.993 & 2.071 & 2.103 \\
    $\gamma_{r,0}$ (meV)  & 3.869 & 0.988 & 0.265  & 6.426  \\
    $\gamma_{nr}$ (meV)   & 11.340 & 12.260 & 12.070 & 45.09 \\
  \end{tabular}
  \end{ruledtabular}
  \label{tab:decaytable}
\end{table}

\label{sec:additionaldevices}
 
\begin{figure}
    \centering
    \includegraphics[width=0.75\linewidth]{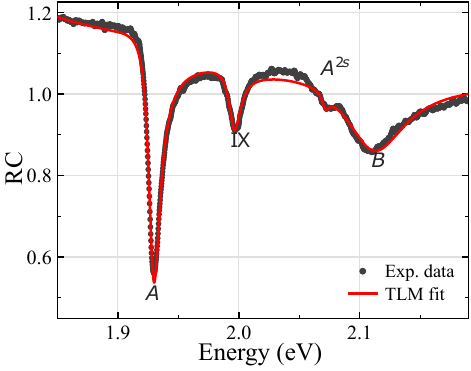}
    \caption{RC measurement from the second reference device at $T = 5~\mathrm{K}$, together with the TLM fit.}
    \label{fig:ref1RC}
\end{figure}
 
\begin{table} 
  \centering
  \caption{Excitonic parameters at $T = 5~\mathrm{K}$ extracted from the second
           SiO$_2$/Si reference device with $\chi_{\mathrm{bg}} = 20.67$.}
  \begin{ruledtabular}
  \begin{tabular}{ldddd}
    & \multicolumn{1}{c}{A}
    & \multicolumn{1}{c}{IX}
    & \multicolumn{1}{c}{A$^{2s}$}
    & \multicolumn{1}{c}{B} \\
    \hline
    $E_0$ (eV)            & 1.928 & 1.996 & 2.071 & 2.109 \\
    $\gamma_{r,0}$ (meV)  & 3.891 & 0.980 & 0.176 & 7.725 \\
    $\gamma_{nr}$ (meV)   & 9.905 & 10.996 & 7.200 & 57.635 \\
  \end{tabular}
  \end{ruledtabular}
  \label{tab:decaytable_ref1}
\end{table}
 
The second device differs in encapsulation thickness, with top and bottom hBN of $48~\mathrm{nm}$ and $26~\mathrm{nm}$ on a $270~\mathrm{nm}$ oxide, compared with $23~\mathrm{nm}$ and $31~\mathrm{nm}$ on a $285~\mathrm{nm}$ oxide for the device of Fig.~\ref{fig:IXcharacterization}. Despite this difference the extracted oscillator strength ratio is reproduced, with the IX $\gamma_{r,0}$ $\approx 25\%$ of that of the A exciton, and the intrinsic weights $W_{0,A} \approx 0.393$ and $W_{0,IX} \approx 0.089$ again place both resonances deep in the undercoupled regime.

\section{Interlayer excitons in trilayer MoS$_2$}
\label{sec:trilayerIX}

\begin{figure}
  \centering
  \includegraphics[width=0.75\linewidth]{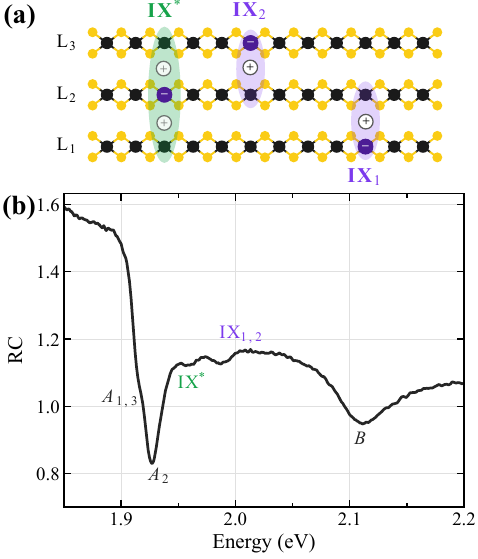}
  \caption{Interlayer excitons in trilayer MoS$_2$.
  (a) Side view of 2H trilayer MoS$_2$ with layers L$_1$ to L$_3$.
  The interlayer excitons IX$_1$ and IX$_2$ have the electron in an outer layer and the hole shared with middle layer L$_2$.
  The quadrupolar exciton IX$^{*}$ has the electron in L$_2$ and the hole shared across all three layers.
  (b) Reflectance contrast RC $=R/R_0$ of the hBN encapsulated trilayer on SiO$_2$/Si at 5 K.
  The intralayer $A$ and $B$ excitons dominate the spectrum.
  The $A$ resonance is a doublet with $A_{1,3}$ at 1.916 eV from the outer layers and $A_{2}$ at 1.925 eV from the middle layer, split by the different exciton binding energy in inner and outer layers.}
  \label{figTrilayerIX}
\end{figure}

Figure~\ref{figTrilayerIX}(a) shows the different interlayer exciton configurations of trilayer MoS$_2$.
In IX$_1$ and IX$_2$ the electron sits in an outer layer and the hole is shared with the middle layer L$_2$.
The two dipoles point in opposite directions and the two states are degenerate, so they appear as one resonance IX$_{1,2}$.
In IX$^{*}$ the electron sits in L$_2$ and the hole is shared across all three layers \cite{Leisgang2020GiantMoS2}.
 
Figure~\ref{figTrilayerIX}(b) shows the 5 K reflectance contrast of the hBN encapsulated trilayer on SiO$_2$/Si.
The $A$ resonance is a doublet, with $A_{1,3}$ at 1.916 eV from the two outer layers and $A_{2}$ at 1.925 eV from the middle layer, split by their different exciton binding energy \cite{Leisgang2020GiantMoS2}.
IX$^{*}$ appears at 1.958 eV and IX$_{1,2}$ at 1.989 eV, close to the bilayer IX at 1.996 eV, and the $B$ exciton lies at 2.109 eV.
Both interlayer features are weak on SiO$_2$/Si because of their small $\gamma_{r,0}$ and large  $\gamma_{nr}$.

\section{Transmission-line model}
\label{sec:tlm}
We derive $P$, $Q$ and $\zeta$ of Sec.~\ref{sec:tlm_main} for the
general stack
\begin{equation}
  \begin{split}
    \text{air}\,/\,&\text{superstrate}\,(n_1,d_1)\,/\,\text{hBN}\,(d_t)\,/\\[2pt]
    &\text{TMD}\,(\sigma)\,/\,\text{hBN}\,(d_b)\,/\,\text{substrate}\,(n_S),
  \end{split}
  \label{eq:structure}
\end{equation}
from which every geometry of Table~\ref{tab:cases} follows through its
own terminations. Conventions follow
Ref.~\cite{Eini2025TransmissionHeterostructures}.
\begin{table} 
  \centering
  \caption{Photonic environments treated in this work. All are described
  by the same three admittances evaluated at the bilayer plane,
  $Y_{\mathrm{below}}$, $Y_{\mathrm{above}}$ and
  $\widetilde{Y}_{\mathrm{above}}$, and differ only in what terminates
  the stack above and below.}
  \label{tab:cases}
  \begin{ruledtabular}
  \begin{tabular}{lll}
    Geometry & superstrate & substrate \\
    \hline
    Reference (Fig.~\ref{fig:IXcharacterization})     & absent      & SiO$_2$/Si \\
    Open cavity (Fig.~\ref{fig:fig_sm}) & absent      & Au \\
    Closed cavity (Fig.~\ref{fig:figdm}) & Au, $d_g$   & Au \\
  \end{tabular}
  \end{ruledtabular}
\end{table}
\subsection{Conventions}
The time dependence is $e^{-i\omega t}$ and $z$ increases downward, from
the air into the substrate. At any plane the admittance is the ratio of
the tangential fields, $Y = H_T/E_T$. At normal incidence a medium of
refractive index $n_j$ has characteristic admittance
$Y_j = n_j\varepsilon_0 c$, so that $Y_A = \varepsilon_0 c$ for air,
$Y_1 = n_1\varepsilon_0 c$ for the superstrate,
$Y_{hBN} = n_H\varepsilon_0 c$ for hBN and $Y_S = n_S\varepsilon_0 c$ for
the substrate, with $n_1$ and $n_S$ complex wherever those layers
absorb. The phase accumulated across a layer is $\phi_j = k_0 n_j d_j$
with $k_0 = \omega/c$, and $k_{z,j} = k_0 n_j$ is the propagation
constant of layer $j$ along $z$. The sheet conductivity $\sigma$ is
defined in Eq.~\eqref{eq:sigma}. For the gold devices the substrate is
the back mirror, $Y_S = Y_{\mathrm{Au}}$, and for the closed cavity the
superstrate is the top gold film, $Y_1 = Y_{\mathrm{Au}}$.
 
Across a layer of index $n_j$ and phase $\phi_j$ the admittance
transforms as \cite{Eini2025TransmissionHeterostructures,Kahlon2025ImportanceHeterostructures}
\begin{equation}
  Y = Y_j\,\frac{Y_{L} - iY_j\tan\phi_j}{Y_j - iY_{L}\tan\phi_j},
  \label{eq:transform}
\end{equation}
with $Y_L$ the admittance at the far face of the layer and $Y$ the
admittance at the near face. It includes all internal reflections and
is applied once per layer, from each outer medium toward the bilayer
plane.
\subsection{Effective admittance below the TMD sheet}
The semi-infinite substrate presents $Y_S$, and one application of
Eq.~\eqref{eq:transform} across the bottom hBN gives
\begin{equation}
  Y_{\mathrm{below}} = Y_{hBN}\,
    \frac{Y_S - iY_{hBN}\tan\phi_b}{Y_{hBN} - iY_S\tan\phi_b} .
  \label{eq:Ybelow}
\end{equation}
For the SiO$_2$/Si reference, Eq.~\eqref{eq:transform} is first applied
across the oxide. The $100~\mathrm{nm}$ gold mirror is effectively
semi-infinite.
\subsection{Effective admittance above the TMD sheet}
Carrying $Y_A$ of the air down through the superstrate and the top hBN
gives
\begin{equation}
  \begin{split}
    Y_{\mathrm{sup}} &= Y_{1}\,
      \frac{Y_A - iY_{1}\tan\phi_1}{Y_{1} - iY_A\tan\phi_1},\\[6pt]
    Y_{\mathrm{above}} &= Y_{\mathrm{hBN}}\,
      \frac{Y_{\mathrm{sup}} - iY_{\mathrm{hBN}}\tan\phi_t}
           {Y_{\mathrm{hBN}} - iY_{\mathrm{sup}}\tan\phi_t},
  \end{split}
  \label{eq:Yabove}
\end{equation}
and the same two steps with every tangent reversed give
\begin{equation}
    \begin{split}
  \widetilde{Y}_{\mathrm{sup}} &= Y_{1}\,
    \frac{Y_A + iY_{1}\tan\phi_1}{Y_{1} + iY_A\tan\phi_1},\\[6pt]
  \widetilde{Y}_{\mathrm{above}} &= Y_{hBN}\,
    \frac{\widetilde{Y}_{\mathrm{sup}} + iY_{hBN}\tan\phi_t}
         {Y_{hBN} + i\widetilde{Y}_{\mathrm{sup}}\tan\phi_t} .
  \label{eq:Yabove_tilde}
  \end{split}
\end{equation}
$Y_{\mathrm{above}}$ describes the outgoing wave, with
$H_T/E_T = -Y_{\mathrm{above}}$ at the bilayer plane, the sign arising
because looking up is along $-z$. $\widetilde{Y}_{\mathrm{above}}$
describes the incoming wave, with
$H_T/E_T = +\widetilde{Y}_{\mathrm{above}}$. When every layer above the
sheet is lossless,
$\widetilde{Y}_{\mathrm{above}} = Y_{\mathrm{above}}^{\,*}$, while an
absorbing superstrate makes the two independent.
\subsection{The sheet}
The bilayer is a surface current $\mathbf{K} = \sigma\mathbf{E}_T$,
across which $E_T$ is continuous and $H_T$ jumps by $\sigma E_T$. The
admittance just above it, looking down, is therefore
\begin{equation}
  Y_{\mathrm{sh}} = Y_{\mathrm{below}} + \sigma .
  \label{eq:shunt}
\end{equation}
\subsection{The reflection amplitude}
In the air there are two waves, one of tangential electric amplitude
$E_{\mathrm{inc}}$ travelling down and one of amplitude
$E_{\mathrm{ref}}$ travelling up, and the reflection amplitude is
$r = E_{\mathrm{ref}}/E_{\mathrm{inc}}$. Since the upper stack is
linear, the fields at the bilayer plane are
\begin{equation}
\begin{split}
  E_T &= \alpha E_{\mathrm{inc}} + \beta E_{\mathrm{ref}} \\[6pt]
  H_T &= \widetilde{Y}_{\mathrm{above}}\,\alpha E_{\mathrm{inc}}
        - Y_{\mathrm{above}}\,\beta E_{\mathrm{ref}} ,
\end{split}
  \label{eq:superpose}
\end{equation}
where $\alpha$ and $\beta$ include all reflections in the upper stack
and depend on neither the sheet nor the substrate. Imposing
$H_T = Y_{\mathrm{sh}} E_T$ from Eq.~\eqref{eq:shunt} gives
\begin{equation}
  \widetilde{Y}_{\mathrm{above}}\,\alpha E_{\mathrm{inc}}
   - Y_{\mathrm{above}}\,\beta E_{\mathrm{ref}}
   = Y_{\mathrm{sh}}\bigl(\alpha E_{\mathrm{inc}}
     + \beta E_{\mathrm{ref}}\bigr) ,
  \label{eq:match}
\end{equation}
and solving for $r$, with $\beta \neq 0$,
\begin{equation}
  r = \frac{\alpha}{\beta}\cdot
      \frac{\widetilde{Y}_{\mathrm{above}} - Y_{\mathrm{below}} - \sigma}
           {Y_{\mathrm{above}} + Y_{\mathrm{below}} + \sigma} .
  \label{eq:r_super}
\end{equation}
\subsection{The RC}
Since $\alpha/\beta$ contains no $\sigma$, it cancels in the ratio to
the bare-cavity amplitude $r_0 = r(\sigma = 0)$,
\begin{equation}
\begin{split}
  \frac{r(\omega)}{r_0}
   &= \frac{\widetilde{Y}_{\mathrm{above}}-Y_{\mathrm{below}}-\sigma}
          {\widetilde{Y}_{\mathrm{above}}-Y_{\mathrm{below}}}
     \cdot
     \frac{Y_{\mathrm{above}}+Y_{\mathrm{below}}}
          {Y_{\mathrm{above}}+Y_{\mathrm{below}}+\sigma}\\[4pt]
   &= \frac{1-\sigma Q}{1+\sigma P} \equiv 1-Z
\end{split}
  \label{eq:Zratio}
\end{equation}
with $Z = (P+Q)/(\sigma^{-1}+P)$ and
\begin{equation}
  P = \frac{1}{Y_{\mathrm{above}} + Y_{\mathrm{below}}},
  \qquad
  Q = \frac{1}{\widetilde{Y}_{\mathrm{above}} - Y_{\mathrm{below}}} ,
  \label{eq:PQ_supp}
\end{equation}
which are Eq.~\eqref{eq:master} of Sec.~\ref{sec:tlm_main}. The
reflection contrast follows as
\begin{equation}
  \frac{R(\omega)}{R_0(\omega)} = \left|1-Z\right|^2
   = 1 - 2\,\mathrm{Re}\,Z + |Z|^2 .
  \label{eq:RCdef}
\end{equation}
$P$ and $Q$ can thus be interpreted as the impedances seen at the
bilayer plane by the sheet and by the incident wave respectively.
 
\subsection{Closed forms and the cavity function}
Let $n_u$ denote the effective index terminating the stack above the top
hBN, equal to $1$ when the superstrate is absent and to
$n_1(1 - i n_1\tan\phi_1)/(n_1 - i\tan\phi_1)$ otherwise, with
$\widetilde n_u$ the same quantity with $\tan\phi_1$ negated. The
round-trip reflection coefficients referred to the bilayer plane are
\begin{equation}
  \begin{split}
    r_{b} &= \frac{n_{H}-n_S}{n_{H}+n_S}\,e^{2i\phi_b},\\[4pt]
    r_{t} &= \frac{n_{H}-n_u}{n_{H}+n_u}\,e^{2i\phi_t},\\[4pt]
    \widetilde{r}_{t} &= \frac{n_{H}-\widetilde n_u}
                             {n_{H}+\widetilde n_u}\,e^{-2i\phi_t},
  \end{split}
  \label{eq:rbrt}
\end{equation}
with arguments
\begin{equation}
  \psi_b = \arg r_b,
  \qquad
  \psi_t = \arg r_t .
  \label{eq:rhopsi}
\end{equation}
For a bare hBN/air top interface the modulus takes the value
\begin{equation}
  r_{HA} \equiv \frac{n_{H}-1}{n_{H}+1} \approx 0.38 ,
  \label{eq:rHA}
\end{equation}
which is the ceiling on $|r_t|$ in the absence of a top mirror. Each
chain then reduces to its round-trip coefficient,
\begin{equation}
  \begin{split}
    Y_{\mathrm{above}} &= \varepsilon_0 c\,n_{H}\,\frac{1-r_{t}}{1+r_{t}},\\[4pt]
    Y_{\mathrm{below}} &= \varepsilon_0 c\,n_{H}\,\frac{1-r_{b}}{1+r_{b}},\\[4pt]
    \widetilde{Y}_{\mathrm{above}} &= \varepsilon_0 c\,n_{H}\,
        \frac{1-\widetilde{r}_{t}}{1+\widetilde{r}_{t}},
  \end{split}
  \label{eq:Y_in_r}
\end{equation}
and Eq.~\eqref{eq:PQ_supp} becomes
\begin{equation}
  \frac{1}{\varepsilon_0 c\,P}
  = \frac{2n_{H}\,(1-r_{b}r_{t})}{(1+r_{b})(1+r_{t})},
  \qquad
  \frac{1}{\varepsilon_0 c\,Q}
  = \frac{2n_{H}\,(r_{b}-\widetilde{r}_{t})}
         {(1+\widetilde{r}_{t})(1+r_{b})}.
  \label{eq:PQ_closed}
\end{equation}
The first fixes the cavity impedance through the dimensionless cavity
function
\begin{equation}
  \mathcal{F}(\omega) \equiv 2\varepsilon_0 c\,n_{H}\,P
  = \frac{(1+r_{b})(1+r_{t})}{1 - r_{b}r_{t}},
  \label{eq:calF_supp}
\end{equation}
which is Eq.~\eqref{eq:calF}. When the layers above the sheet are
lossless, $\widetilde{r}_t = r_t^{\,*}$ and $Q$ and $P$ are not
independent.
\subsection{Cavity-dressed lineshape and the Purcell factor}
\label{sec:lamb}
Split the sheet conductivity into a non-resonant background and one
Lorentzian drawn from Eq.~\eqref{eq:chi},
\begin{equation}
  \sigma(\omega) = \sigma_{\mathrm{bg}} + \sigma_x ,
  \qquad
  \sigma_x = \frac{i\,\varepsilon_0 c\,\gamma_{r,0} (\omega/\omega_0)}
                  {\omega-\omega_0+i\gamma_{nr}/2} ,
  \label{eq:sigma_split}
\end{equation}
where the other oscillators and $\chi_{\mathrm{bg}}$ are absorbed into
$\sigma_{\mathrm{bg}}$. Equation~\eqref{eq:Zratio} then factorizes as
\begin{equation}
\begin{gathered}
  1 - Z = \bigl(1-Z_{\mathrm{bg}}\bigr)\bigl(1-Z_x\bigr),\\[4pt]
  Z_{\mathrm{bg}} = \frac{\mathcal{L}}{\sigma_{\mathrm{bg}}^{-1}+P},
  \qquad
  Z_x = \frac{\mathcal{L}'}{\sigma_x^{-1}+P'},
\end{gathered}
  \label{eq:pullout}
\end{equation}
with $\mathcal{L} = P+Q$, $\mathcal{L}' = P'+Q'$ and the renormalized
impedances
\begin{equation}
  \frac{1}{P'} = \frac{1}{P} + \sigma_{\mathrm{bg}},
  \qquad
  \frac{1}{Q'} = \frac{1}{Q} - \sigma_{\mathrm{bg}} .
  \label{eq:PQprime}
\end{equation}
Each factor has the form of Eq.~\eqref{eq:master}, the background
entering only as a shunt on $1/P$. $Z_{\mathrm{bg}}$ is smooth across a
linewidth, and inserting $\sigma_x$ into $Z_x$ gives a single Lorentzian
\begin{equation}
  Z_x(\omega) = \frac{i\,\varepsilon_0 c\,\gamma_{r,0}\,(\omega/\omega_0)\,\mathcal{L}'}
                     {(\omega-\widetilde\omega)+i\,\Gamma_T/2},
  \label{eq:lineshape}
\end{equation}
with the cavity-renormalized resonance frequency and total width
\begin{equation}
\begin{gathered}
  \widetilde\omega = \omega_0
    + \varepsilon_0 c\,\gamma_{r,0}\,\mathrm{Im}\,P' ,\\[4pt]
  \Gamma_T = \gamma_{nr} + \zeta\,\gamma_{r,0},
  \qquad
  \zeta = 2\varepsilon_0 c\,\mathrm{Re}\,P' ,
\end{gathered}
  \label{eq:GammaT_supp}
\end{equation}
which is Eq.~\eqref{eq:GammaT}. Only $\gamma_r$ is dressed, and
$\mathrm{Im}\,P'$ gives a cavity induced resonance shift. Both are evaluated at the
pole, where $\omega/\omega_0$ differs from unity by about $10^{-3}$ and
the mirror dispersion of $P'$ is negligible across a linewidth.
 
Combining Eqs.~\eqref{eq:PQprime} and \eqref{eq:calF_supp},
\begin{equation}
  \zeta(\omega_0)
  = \mathrm{Re}\!\left[\frac{\mathcal{F}(\omega_0)}
     {n_H + \dfrac{\sigma_{\mathrm{bg}}}{2\varepsilon_0 c}\,
      \mathcal{F}(\omega_0)}\right],
  \label{eq:zeta_supp}
\end{equation}
which is Eq.~\eqref{eq:zeta}. It contains no decay rates, which is the
basis of the design rule of Sec.~\ref{sec:opencavities}.
\subsection{Bounds on the Purcell factor}
\label{sec:bounds}
Neglecting $\sigma_{\mathrm{bg}}$, which as a pure susceptance only
detunes the cavity, Eq.~\eqref{eq:zeta_supp} gives
$\zeta = \mathrm{Re}\,\mathcal{F}/n_H$. The field factors $1+r_b$ and
$1+r_t$ of Eq.~\eqref{eq:calF_supp} are largest at
$\psi_b = \psi_t = 0$, and the denominator $1-r_br_t$ is smallest
at $\psi_b+\psi_t = 0$. Both hold at $\psi_b = \psi_t = 0$, where the
sheet sits at an antinode, while $\psi_b = \psi_t = \pi$ is the node.
Setting $r_j = \pm|r_j|$ gives
\begin{equation}
  \zeta_{\max} = \frac{(1+|r_b|)(1+|r_t|)}{n_{H}(1-|r_b||r_t|)},
  \qquad
  \zeta_{\min} = \frac{(1-|r_b|)(1-|r_t|)}{n_{H}(1-|r_b||r_t|)} .
  \label{eq:zeta_bounds_supp}
\end{equation}
With $|r_b| \to 1$ and $|r_t| = r_{HA}$ of Eq.~\eqref{eq:rHA} they
return the Salisbury-screen value $\zeta_0 = 2$.

\section{Open cavities in the transmission-line model}
\label{sec:opencav_tlm}

For the open cavity the substrate is the opaque $100~\mathrm{nm}$ gold
mirror.
\subsection{Absorption}
Because the transmission through the stack vanishes, the reflection
contrast of Eq.~\eqref{eq:RCdef} gives the absorption directly as
\begin{equation}
  A \equiv 1 - \frac{R(\omega)}{R_0(\omega)}
            = 2\,\mathrm{Re}\,Z(\omega) - |Z(\omega)|^2 .
  \label{eq:Adef}
\end{equation}
\subsection{At-resonance absorption and critical coupling}
\label{sec:resonance}
With Eq.~\eqref{eq:pullout} the absorption becomes
\begin{equation}
  A = 1 - \mathcal{B}\,\bigl|1-Z_x\bigr|^{2}
    = A_{\mathrm{bg}}
      + \mathcal{B}\bigl(2\,\mathrm{Re}\,Z_x - |Z_x|^{2}\bigr),
  \label{eq:A_factorized}
\end{equation}
with the baseline factor
$\mathcal{B} \equiv |1-Z_{\mathrm{bg}}|^{2}$ and the baseline absorption
$A_{\mathrm{bg}} = 1-\mathcal{B}$. Since $\sigma_{\mathrm{bg}}$ is a
pure susceptance and $|\sigma_{\mathrm{bg}}\mathcal{L}| \ll 1$,
$|\mathcal{B}-1| < 10^{-2}$ for all fabricated devices at both
resonances, and we set $\mathcal{B} = 1$.
 
Evaluating Eq.~\eqref{eq:lineshape} at $\omega = \widetilde\omega$ gives
$Z_x = 2\varepsilon_0 c\,\mathcal{L}'X$ with
$X \equiv \gamma_{r,0}/\Gamma_T$, and inserting this into
Eq.~\eqref{eq:A_factorized} at $\mathcal{B} = 1$ yields the
at-resonance peak absorption
\begin{equation}
  A_{\max} = \zeta_1 X\,(1 - \zeta_2 X),
  \qquad
  X = \frac{\gamma_{r,0}/\gamma_{nr}}{1+\zeta\,\gamma_{r,0}/\gamma_{nr}},
  \label{eq:Amax_supp}
\end{equation}
where $X$ is the radiative fraction of the dressed linewidth and
\begin{equation}
  \zeta_1 = 4\varepsilon_0 c\,\mathrm{Re}\,\mathcal{L}',
  \qquad
  \zeta_2 = \frac{\varepsilon_0 c\,|\mathcal{L}'|^2}
                 {\mathrm{Re}\,\mathcal{L}'}
  \label{eq:zeta12}
\end{equation}
are real geometric prefactors. If $\mathrm{Im}\,\mathcal{L}' = 0$ then
$\zeta_1 = 4\zeta_2$, and if also $\mathrm{Re}\,Q' = \mathrm{Re}\,P'$
then $\zeta_2 = \zeta$. Both hold closely for the hBN/Au stack over the
fabricated thickness range, and Eq.~\eqref{eq:Amax_supp} reduces to
\begin{equation}
  A_{\max} = \frac{4\,W}{(1+W)^2},
  \qquad W = \frac{\zeta \gamma_{r,0}}{\gamma_{nr}},
  \label{eq:Amax_cc_supp}
\end{equation}
which is Eq.~\eqref{eq:Amax_perfect}.
\subsection{Monotonicity and the design rule}
\label{sec:monotone}
Differentiating Eq.~\eqref{eq:Amax_cc_supp},
\begin{equation}
  \frac{dA_{\max}}{dW}
   = \frac{4\,(1-W)}{(1+W)^{3}},
  \label{eq:dAmax}
\end{equation}
which is positive for $W<1$. For an undercoupled resonance $A_{\max}$
therefore increases monotonically with $\zeta$, so the geometry that
maximizes $\zeta$ also maximizes the absorption. The decay rates set the
absolute absorption but not the optimal geometry, which by
Eq.~\eqref{eq:zeta_supp} follows from the resonance energy and
$\chi_{\mathrm{bg}}$ alone.

\subsection{Energy dependence of the Purcell factor}
\begin{figure} 
    \centering
    \includegraphics[width=0.75\linewidth]{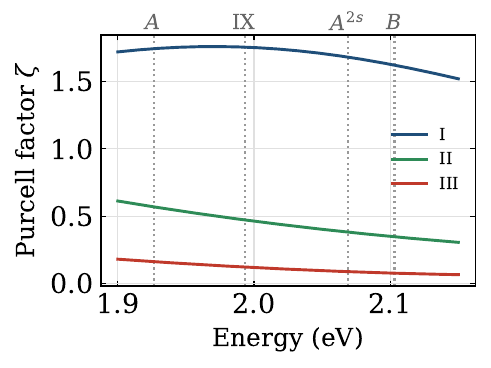}
    \caption{Energy dependence of the Purcell factor for the
    open-cavity devices I, II and III.}
    \label{fig:zeta_SM_energy}
\end{figure}

The Purcell factor for given top and bottom hBN thicknesses is dispersive
because the round-trip hBN phases change with photon energy.
Figure~\ref{fig:zeta_SM_energy} plots $\zeta(E)$ in the relevant energy
window for the three open-cavity devices I, II and III. In all three the
Purcell factor shows weak dispersion.

\subsection{Temperature-dependent measurements}
\label{sec:temperature}
\begin{figure} 
    \centering
    \includegraphics[width=0.75\linewidth]{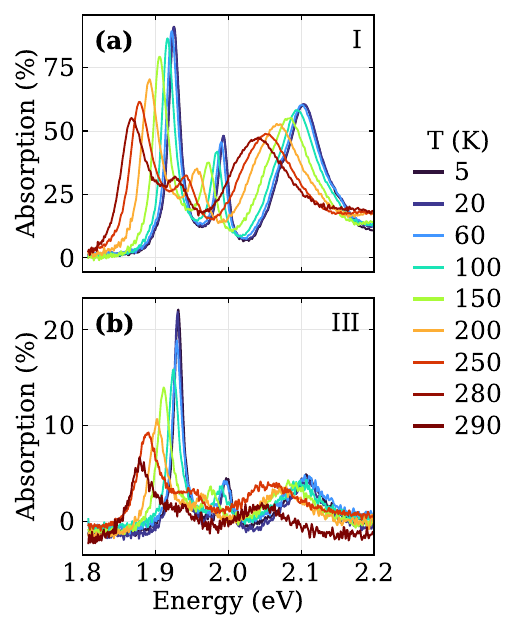}
    \caption{Temperature-dependent absorption of the open-cavity devices
    I~(a) and III~(b) from $5$ to $290~\mathrm{K}$. The Purcell enhancement
    of device~I and the suppression of device~III both persist to room
    temperature.}
    \label{fig:TD_SM}
\end{figure}

Figure~\ref{fig:TD_SM} shows the absorption of the open-cavity devices
from $5$ to $290~\mathrm{K}$. The IX absorption of device~I stays enhanced
and that of device~III stays suppressed over the whole range. All
resonances broaden as the nonradiative rate rises with temperature and
both devices are driven further into the undercoupled regime. The
enhancement reported in Sec.~\ref{sec:opencavities} is therefore robust
to room temperature.

\subsection{Photoluminescence}
\begin{figure}
    \centering
    \includegraphics[width=0.75\linewidth]{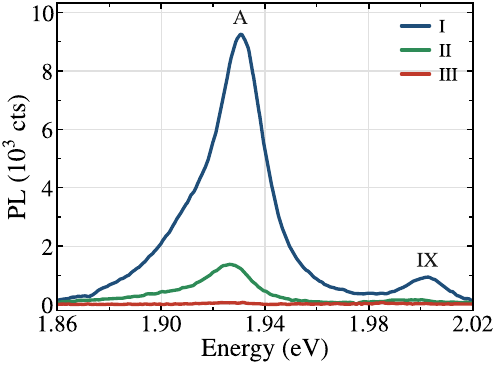}
    \caption{Photoluminescence spectra of devices I, II and III captured under identical excitation conditions,
    showing that the cavity-enhanced absorption in device I is accompanied
    by substantially brighter IX emission.}
    \label{fig:pl}
\end{figure}

Photoluminescence was measured on the same devices in the same setup under
$532~\mathrm{nm}$ continuous-wave excitation at $1~\mu\mathrm{W}$
(Appendix~\ref{sec:measurement}). All spectra were acquired at
$T = 5~\mathrm{K}$ under identical excitation and collection conditions so
that the three devices can be compared directly.
Figure~\ref{fig:pl} shows the result. The emission of A and IX in device I
is significantly brighter than in devices II and III.

\section{Closed cavity}
\label{sec:dm_appendix}
\subsection{Dependence on the top gold thickness}
\begin{figure} 
    \centering
    \includegraphics[width=0.75\linewidth]{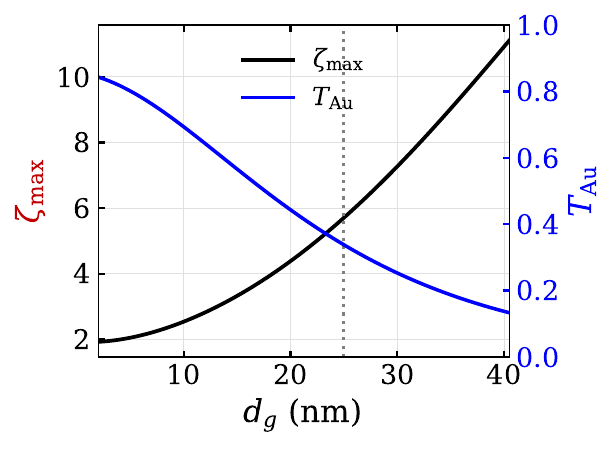}
    \caption{Maximum accessible Purcell factor $\zeta_{\max}$ (left axis)
    and the transmittance into the cavity $T_{\mathrm{Au}}$ (right axis)
    versus top gold thickness $d_g$, evaluated at the IX energy.
    $T_{\mathrm{Au}}$ is the transmittance of the air/Au/hBN top
    interface, i.e.\ the fraction of incident power coupled through the
    top mirror into the cavity. }
    \label{fig:dgdep}
\end{figure}

The top gold thickness $d_g$ sets the trade-off between reflectivity $|r_t|$ and hence the maximum accessible Purcell factor
and transmission into the cavity. Figure~\ref{fig:dgdep} shows the maximum accessible
Purcell factor $\zeta_{\max}(d_g)$ and the transmittance into the cavity
$T_{\mathrm{Au}}(d_g)$ at the IX energy. Increasing $d_g$ raises the
effective top reflectance and hence $\zeta_{\max}$, but exponentially
suppresses the power coupled into the cavity. The fabricated value
$d_g = 25~\mathrm{nm}$ balances the two, reaching
$\zeta_{\max}\approx 5.7$ while retaining $T_{\mathrm{Au}}\approx 34\%$.

\subsection{Transmission-line fit of the closed cavity}
\label{sec:dm_fit}
The closed cavity is described by the TLM of Appendix~\ref{sec:tlm} with
the top gold film as superstrate, $Y_1 = Y_{\mathrm{Au}}$, and the gold
back mirror as substrate. The bilayer enters as a sheet of physical
thickness $d_{\mathrm{TMD}} = 1.4~\mathrm{nm}$ placed between the two
hBN layers, so the device stack is $d_t + d_{\mathrm{TMD}} + d_b$ thick
while the adjacent bare cavity contains the hBN alone. Both measured
spectra are normalized by the reflectance of the gold substrate recorded
in the same run.

\begin{figure} 
  \centering
  \includegraphics[width=0.75\linewidth]{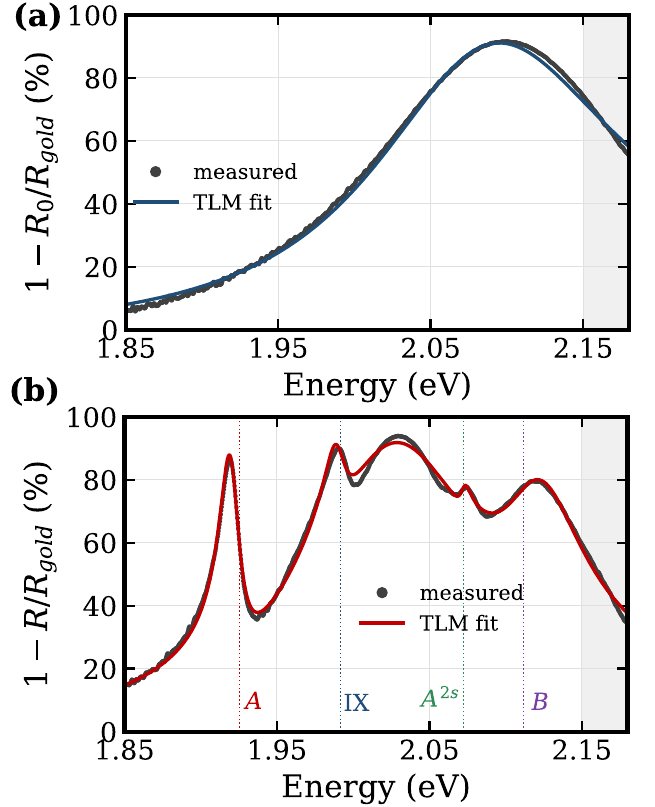}
  \caption{Transmission-line fits of the closed cavity at
    $T = 5~\mathrm{K}$, shown as the absorption of the stack
    $1-R/R_{\mathrm{Gold}}$. (a)~Bare cavity fitted at $\sigma = 0$,
    which fixes $d_g$ and the total hBN thickness $D$. (b)~Device with
    the bilayer, fitted with $d_g$ and $D$ frozen,
    $d_t = 48~\mathrm{nm}$, $d_b = 23~\mathrm{nm}$ and
    $d_{\mathrm{TMD}} = 1.4~\mathrm{nm}$.}
  \label{fig:dm_fits}
\end{figure}

The fit proceeds in two stages so that the cavity is characterized
before the bilayer is introduced. In the first stage the bare-cavity
spectrum $1-R_0/R_{Gold}$, is fitted at $\sigma = 0$ [Fig.~\ref{fig:dm_fits}(a)]. Without
the sheet the response depends only on the gold thickness ($d_g$) and on the
total hBN thickness $D = d_t + d_b$. The fit returns $d_g = 24.9~\mathrm{nm}$ and
$D = 71.0~\mathrm{nm}$, in agreement with the deposition and AFM values,
and reproduces the Fabry-P\'erot mode at $2.10~\mathrm{eV}$. These two
parameters are held fixed from here on.

In the second stage the device spectrum $R$ is fitted with
$d_t = 48~\mathrm{nm}$ and $d_b = 23~\mathrm{nm}$
[Fig.~\ref{fig:dm_fits}(b)]. The resulting parameters are listed in
Table~\ref{tab:dmfit}. The fit gives $\chi_{bg} = 21.5$, and radiative
rates of $3.60~\mathrm{meV}$ for the A exciton and $0.96~\mathrm{meV}$
for the IX similar to that  of the reference device of
Table~\ref{tab:decaytable}.

\begin{table} 
  \centering
  \caption{Excitonic parameters of the closed cavity at
           $5~\mathrm{K}$ from the fit of Fig.~\ref{fig:dm_fits}(b) with
           $\chi_{bg} = 21.5$.}
  \label{tab:dmfit}
  \begin{ruledtabular}
  \begin{tabular}{ldddd}
    & \multicolumn{1}{c}{A}
    & \multicolumn{1}{c}{IX}
    & \multicolumn{1}{c}{A$^{2s}$}
    & \multicolumn{1}{c}{B} \\
    \hline
    $E_0$ (eV)            & 1.9256 & 1.9916 & 2.0725 & 2.1119 \\
    $\gamma_{r,0}$ (meV)  & 3.602  & 0.957  & 0.163  & 6.072  \\
    $\gamma_{nr}$ (meV)   & 10.41  & 13.29  & 7.62   & 47.64  \\
  \end{tabular}
  \end{ruledtabular}
\end{table}

\clearpage
\bibliography{reference_corrected}
\end{document}